\documentclass[%
 reprint,
superscriptaddress,
 amsmath,amssymb,
 aps,
]{revtex4-2}

\usepackage{graphicx}
\usepackage{dcolumn}
\usepackage{color}
\usepackage{multirow}
\usepackage{comment}
\usepackage{algpseudocode}
\usepackage{algorithm}
\usepackage{bm}
\usepackage{mathrsfs}
\usepackage{float}
\graphicspath{{Images/}}

\usepackage{xcolor}

	\usepackage[T1]{fontenc}
	\usepackage[utf8]{inputenc}
	\usepackage[english]{babel}

\usepackage{hyperref}
\definecolor{red1}{rgb}{0.6,0,0}

\begin{document}

\title{Metaheuristic conditional neural network for harvesting skyrmionic metastable states}

\newcommand{\KTH}{Department of Applied Physics, School of Engineering Sciences, KTH Royal Institute of Technology, 
AlbaNova University Center, SE-10691 Stockholm, Sweden}

\newcommand{\KTHeecs}{Division of Computational Science and Technology, School of Electrical Engineering and Computer Science, KTH Royal Institute of Technology, AlbaNova University Center, SE-10691 Stockholm, Sweden}

\newcommand{\SeRC}{SeRC (Swedish e-Science Research Center), KTH Royal Institute of Technology, SE-10044 Stockholm, Sweden}
\newcommand{\Uppsala}{Department of Physics and Astronomy, Uppsala University, Box 516, SE-75120 Uppsala, Sweden}
\newcommand{\Orebro}{School of Science and Technology, \"Orebro University, SE-701 82, \"Orebro, Sweden}
\newcommand{\Stockholm}{Department of Materials and Environmental Chemistry, Stockholm University, SE-10691 Stockholm, Sweden}
\newcommand{\UppsalaChem}{Department of Chemistry - Ångström Laboratory, Uppsala University, Box 538, Uppsala, SE-751 21, Sweden}

\newcommand{\LU}{Department of Physics and Electrical Engineering, Linnaeus University, Hus Magna, SE-39231 Kalmar, Sweden}

\newcommand{\UI}{Science Institute, University of Iceland, 107 Reykjav\'ik, Iceland}

\author{Qichen Xu}
    \affiliation{\KTH}
    \affiliation{\SeRC}
 \author{I. P. Miranda*}
     \affiliation{\Uppsala}
     \thanks{These two authors contributed equally}
\author{Manuel Pereiro*}
    \affiliation{\Uppsala}
    \thanks{These two authors contributed equally}
\author{Filipp N. Rybakov}
     \affiliation{\Uppsala}
\author{Danny Thonig }
     \affiliation{\Orebro}
     \affiliation{\Uppsala}
\author{Erik Sjöqvist}
     \affiliation{\Uppsala}
\author{Pavel F. Bessarab}
     \affiliation{\LU}
     \affiliation{\UI}
\author{Anders Bergman}
     \affiliation{\Uppsala}
\author{Olle Eriksson}
     \affiliation{\Uppsala}     
\author{Pawel Herman}
    \affiliation{\KTHeecs}
    \affiliation{\SeRC}
    
\author{Anna Delin}
    \affiliation{\KTH}
    \affiliation{\SeRC}

\date{\today}

\begin{abstract}
We present a metaheuristic conditional neural-network-based method aimed at identifying physically interesting metastable states in a potential energy surface of high rugosity.
To demonstrate how this method works, we identify and analyze
spin textures with topological charge $Q$ ranging from 1 to $-13$ (where antiskyrmions have $Q<0$) in the Pd/Fe/Ir(111) system, which we model using a classical atomistic spin Hamiltonian based on parameters computed from density functional theory. 
To facilitate the harvest of relevant spin textures, we make use of the newly developed Segment Anything Model (SAM).
Spin textures with  $Q$ ranging from  $-3$ to $-6$ are further analyzed using finite-temperature spin-dynamics simulations. We observe that for temperatures up to around 20\,K, lifetimes longer than 200\,ps are predicted, and that when these textures decay, new topological spin textures are formed. 
We also find that the relative stability of the spin textures depend linearly on the topological charge, but only when comparing the most stable antiskyrmions for each topological charge.
In general, the number of holes (i.e., non-self-intersecting curves that define closed domain walls in the structure) in the spin texture is an important predictor of stability -- the more holes, the less stable is the texture. 
Methods for systematic identification and characterization of complex metastable skyrmionic textures -- such as the one demonstrated here -- are highly relevant for advancements in the field of topological spintronics.
\end{abstract}

\maketitle

\section{Introduction}
In systems with many interacting entities, e.g., a chain of amino acids forming a protein \cite{protein_PES}, a metal cluster \cite{alloys}, or a system of spins \cite{bessarab2018lifetime}, the Hamiltonian and its corresponding potential energy surface (PES) can quickly become highly convoluted as the number of interacting particles increases. This complexity is expressed in the form of rugosity of the PES, which in the case of coupled spins is driven by frustration or competing spin-spin interactions.
In such cases, finding the global minimum is usually a challenging task due to the non-convexity of the high-dimensional function that represents the PES~\cite{oganov2019structure}. 

However, not only the global minimum, but also the numerous local minima in the PES can provide important information about a given system. 
In fact, many of the states corresponding to local minima can be long-lived at finite temperatures and have significant consequences for the physical behavior of the system. 

Magnetic topological spin textures, e.g. skyrmions, are prime examples of such states. Depending on the intrinsic properties (interactions, geometry, anisotropy) and external conditions (field, temperature), skyrmions can emerge as metastable (excited) states in the configuration space, as confirmed by previous experiments \cite{Herve2018,Meyer2019,Oike2016,Karube2017}.

%
Recently, more complex metastable skyrmionic textures such as skyrmion bags, skyrmionium, and skyrmion bundles  have been observed  experimentally\,\cite{Foster2019,Tang2021natnano,Zhang2018,Hagemeister2018,Wang2022natcom}. From the theory side, it is clear that chiral magnetic skyrmions with arbitrary topological charge should exist \cite{rybakov2019chiral}.
%
%
This current interest in more complex metastable skyrmionic textures stems from that such structures may open the door to new topologically inspired spintronic and information-storage concepts. 
For example, combinations of several different types of skyrmionic structure might prove very useful to make the concept of a skyrmionic bit applicable in practice\,\cite{Gobel2021}.
Another example is skyrmion-based artificial synapses for skyrmion-based neuromorphic computing\,\cite{Song2020}.
In addition, the skyrmion Hall, skyrmion Seebeck, and skyrmion Nernst effects depend strongly on the topological charge.\cite{Weissenhofer2022SR}
%

%
From theory and computer simulations, some high-order topological spin textures have been identified\,\cite{Kuchkin2020,Zhang2017natcom,rybakov2019chiral,Leonov2015,Rozsa2017,Carvalho2021}, but a vast flora with novel and potentially very interesting properties remains to be discovered and analyzed.
%
%
This has motivated us to develop a computational method that can systematically harvest metastable skyrmionic spin textures, given a spin Hamiltonian defining the PES.

\begin{figure*}
    \centering
    \includegraphics[width=16cm]{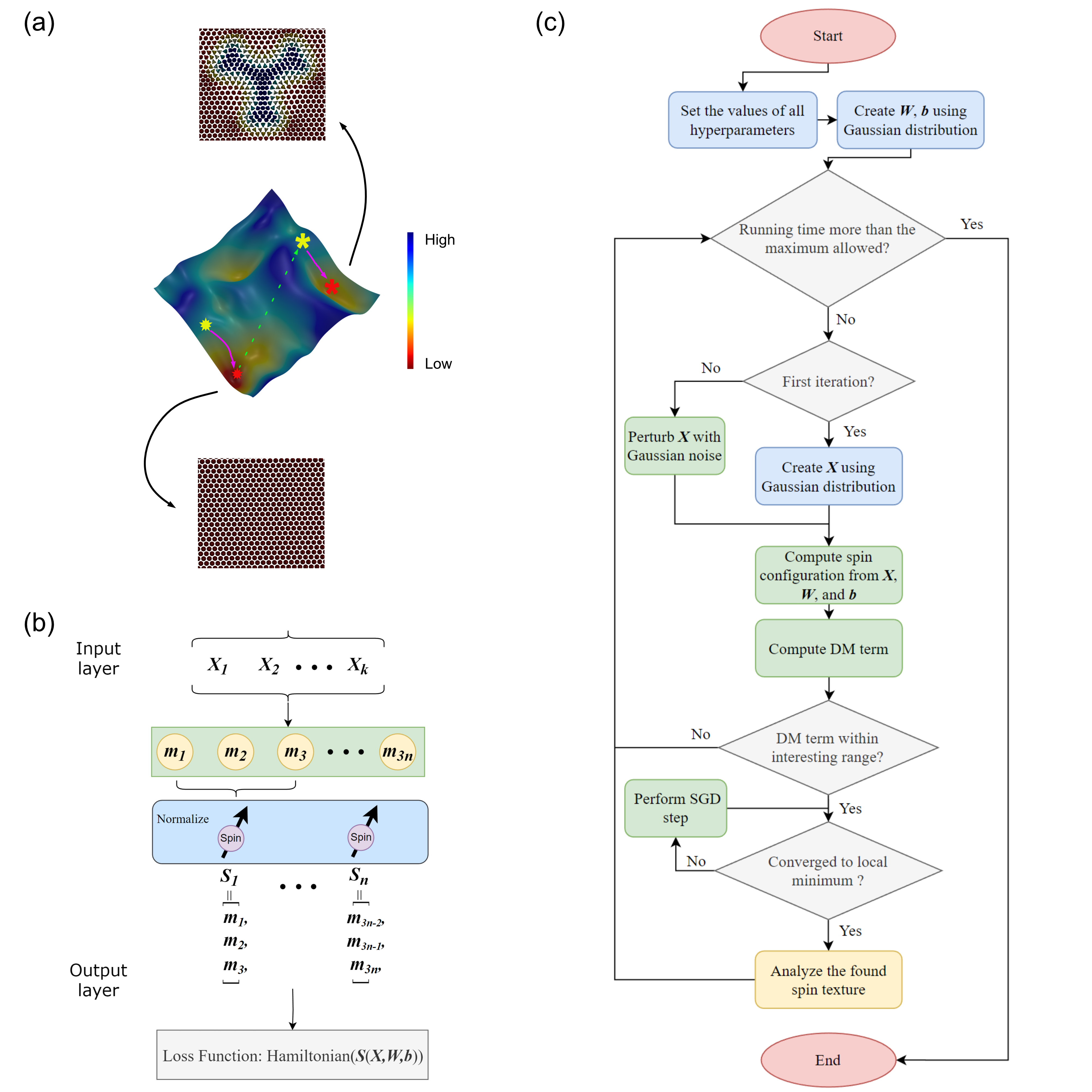}   
    \caption {
(a) 
Conceptual illustration of the overall problem. In a complex PES (the middle picture), we wish to efficently identify local minima containing interesting skyrmionic spin textures (top picture) and avoid converging toward minima with uninteresting spin textures (bottom picture). The yellow and red stars show the initial and final positions for the corresponding NN-SGD optimizations.
(b) 
Chart of the fully connected feed-forward NN-SGD. A random vector $X$ of length $k$ is used to select the starting point of the SGD optimization. The values $m_i$ on the output nodes define the spin configuration, see Eq.\,\eqref{eq:nn-node}. The loss function $\mathscr{L}$ is the spin Hamiltonian, Eq.\,\eqref{eq:param-hamiltonian}.
(c) 
Flow chart showing the main features of the algorithm. For more details, see Section\,\ref{sec:neural-network}.
\label{fig:workflow}
}
\end{figure*}

Here, we present and use a metaheuristic algorithm based on gradient-descent optimization within a conditional neural network, driven by random perturbations. A conceptual illustration of our method is given in Fig.\,\ref{fig:workflow}(a). The goal of our method is to efficiently explore the rugged PES and its multiple energy minima, and in this way identify metastable and possibly long-lived states represented by the Hamiltonian. 
We demonstrate our method by investigating higher-order antiskyrmions within a model of a well-known frustrated system of interacting spins -- the Pd/Fe/Ir(111) film. This system has been very thoroughly characterized experimentally and was one of the first studied systems containing two-dimensional magnetic skyrmions~\cite{romming2013writing}, and has somewhat attained the status of "fruit fly" in the research on topological spin textures. 

\section{Method}
\label{method}
\subsection{Metaheuristic conditional neural network}
\label{sec:neural-network}
Our method consists essentially of two main ingredients: stochastic gradient-descent (SGD) optimization of the Hamiltonian parameterized in a shallow neural network framework, 
and a condition directing the global exploration of the PES.
Both are described in more detail below.

For the first ingredient -- i.e., the shallow neural network, in the following called NN-SGD, whose role is to identify local minima of the Hamiltonian -- we use the AdamW 
\cite{loshchilov2017decoupled,paszke2019pytorch}
implementation of SGD within a single-layer feedforward NN to minimize a loss function $\mathscr{L}(X,W,b)$. Here, $X$ is an input vector defining the starting point of the SGD optimization, whereas $W$ and $b$ are the weights and biases within the NN.
The value of the loss function $\mathscr{L}(X,W,b)$ is provided by the Hamiltonian, i.e., the total energy of the system. 
We use a Heisenberg-type classical atomistic spin Hamiltonian of the form
\begin{equation}\label{eq:param-hamiltonian}
\begin{aligned}
\mathscr{L}(X,W,b)
&=\mathscr{H}(\mathbf{S}_{1}, \mathbf{S}_{2}, \ldots,\mathbf{S}_{n}) =\\
&-
\sum_{i \neq j} J_{i j} \mathbf{S}_{i} \cdot \mathbf{S}_{j}-
\sum_{i \neq j} \mathbf{D}_{i j}\cdot\left(\mathbf{S}_{i} \times \mathbf{S}_{j}\right) \\
&- 
\sum_{i} \mu_{i}\mathbf{B}^{\mathrm{ext}} \cdot \mathbf{S}_{i}  - \sum_{i}K_i^{\mathrm{U}}\left(\mathbf{S}_{i} \cdot \mathbf{e}_{z}\right)^{2},
\end{aligned}
\end{equation}
where $\mathbf{S}_{i}$ is the normalized spin moment at site $i$ and the total number of spins is $n$.  $J_{i j}$, $\mathbf{D}_{i j}$, $K_i^{\mathrm{U}}$, $\mu_{i}$, $\mathbf{e}_{z}$ and $\mathbf{B}^{\mathrm{ext}}$ are Heisenberg exchange interactions, Dzyaloshinskii–Moriya (DM) interactions, uniaxial anisotropy, the magnetic moment length of site $i$, the easy axis vector, and the applied field, respectively.
Thus, to employ SGD for optimization we parameterize the neural network's feed-forward mapping as computation of a spin configuration $(\mathbf{S}_{1}, \mathbf{S}_{2}, \ldots ,\mathbf{S}_{n})$.
The structure of the NN-SGD is illustrated in Fig.\ref{fig:workflow}(b).
To initialize the NN, values for $X$, $W$, and $b$ are randomly assigned from a Gaussian distribution with a mean of 0 and a standard deviation of 1.
The corresponding spin configuration $(S_1, S_2,\ldots,S_n)$ can then be computed from $X$, $W$, and $b$ using
\begin{equation}
\label{eq:nn-node}
\begin{aligned}
M = 
\mathrm{Norm} \left( \mathrm{LeakyReLU} 
\left( X W^T+b \right) 
\right)
\end{aligned}
\end{equation}
where $W^T$ is the transpose of $W$, and $M = (m_1, m_2, m_3, \ldots, m_{3n})$ is a vector containing the components of all spins so that the $x$, $y$, and $z$ components of $\mathbf{S}_{1}$ are $m_1$, $m_2$, and $m_3$, and the components of $\mathbf{S}_{2}$ are $m_4$, $m_5$ and $m_6$, etc. Furthermore, Norm is the normalization operation so that each spin retains unit length,
and LeakyReLU is the activation function, defined in \cite{AW}.
In our implementation, $X$ is a vector of length $k$ and the number of spins $n$, is $10^4$. Typically, $k$ is much smaller than $n$.
%
The role of the NN-SGD is to iteratively converge toward a local minimum of the loss function by adjusting the weights and biases. 
Convergence is achieved when the stopping condition
\begin{equation}\label{eq:Ts}
\begin{aligned}
\left|\mathscr{L}^{(i)}-\mathscr{L}^{(i-1)}\right|<T_s,
\end{aligned}
\end{equation}
is met, i.e., the difference in the loss function between two consecutive checkpoints is small enough. Here, 
$\mathscr{L}^{(i)}$ is the value of the loss function at the $i$-th checkpoint and $T_s$ is the convergence criterion hyperparameter.
In our implementation, typically every 50th optimization step is a checkpoint.


Although the NN-SGD optimisation is conducted with respect to $W$ and $b$, we use an input $X$ to drive the process of the loss function minimisation, as $X$ determines the initial conditions for the SGD and thus it influences which local minimum the NN-SDG converges to. In particular, we use a random vector $X$ and to facilitate "jumping around" over the PES we define new initial conditions by adding random perturbations $P$ to $X$, where $P$ is a vector where some elements are randomly set to zero, and the remaining elements are random numbers taken from a Gaussian distribution. 

This leads us to the second ingredient in our algorithm -- the condition we impose, directing the global exploration of the PES. For each random input vector $X$ generated (i.e., for each new "jump" over the PES), we first compute the value of the 
DM-term $\sum_{i \neq j} \mathbf{D}_{i j}\cdot\left(\mathbf{S}_{i} \times \mathbf{S}_{j}\right)$ in Eq.\,\eqref{eq:param-hamiltonian} 
of the corresponding spin configuration. Only if its value is within a certain predefined range, we proceed with the NN-SGD optimization. This means that we will only investigate local minima pertaining to specific regions of the PES selected by the imposed condition. In more general terms, the approach can be expressed as follows. The imposed condition is
\begin{equation}
\label{HPTP}
\begin{aligned}
T_{\mathrm C}^{\textnormal{low}}<\mathscr{H}_\mathrm C<T_{\mathrm C}^{\textnormal{high}},
\end{aligned}
\end{equation}
where $\mathscr{H}_\mathrm C$
is some in principle arbitrary combination of spin-Hamiltonian terms, and 
$\mathscr{H}_\mathrm C$ as well as the thresholds 
$T_{\mathrm C}^{\textnormal{low}}$  and $T_{\mathrm C}^{\textnormal{high}}$
are selected based on physical insight. 
In practice, the values of $T_{\mathrm C}^{\textnormal{low}}$  and $T_{\mathrm C}^{\textnormal{high}}$ are found through trial and error. 
Also within the SGD loop, we perform occasional checks that the condition Eq.\,\eqref{HPTP} is still valid, in order to minimise the risk of spending the compute time on converging toward an uninteresting local minimum.

In order to sample as many different parts as possible of the PES, we further impose an additional condition 
that the SGD starting point spin configurations we use be dissimilar to each other. In practice, this is done by creating a large number of randomized spin configurations (typically about 200, all fulfilling the condition in Eq.\,\eqref{HPTP}) and, from these, select the one that is most dissimilar to the previously used starting point spin configuration.

When the optimization has converged to a local minimum, the corresponding spin texture is analyzed, for instance by computing its topological charge $Q$. 
The algorithm keeps generating new sets of $X$ and explore the PES as already described above, until the preset maximum execution time is reached. The main parts of the workflow are summarized in a flowchart in Fig.\ref{fig:workflow}(c).

Finally, let us briefly return to the issue of convergence, Eq.\,\eqref{eq:Ts}. We stated previously that the optimization toward the local minimum was performed using SGD. Naturally, this optimization can be done in several ways and there may be methods that are faster than SGD in certain regimes. In this work, we have therefore selected to use not only SGD but also augment it with a non-gradient method based on Metropolis Markov Chain Monte Carlo (MMCMC) when sufficiently close to convergence. 
We will in the following denote the first approach as the \textit{SGD-only mode} 
and the latter approach as the \textit{hybrid mode}.
The SDG-only mode is named this way because it simply uses the SGD inherent in the NN algorithm. In the hybrid mode, we also start out by using SGD, but at a later stage, as convergence is approached, we switch to MMCMC.
In essence, the MMCMC optimizer (as implemented in the Uppsala Atomistic Spin Dynamics, UppASD, package \cite{skubic2008method,UppASD_book}) performs energy minimization under finite temperatures by using the transition probability $P_t$ between two spin configurations in a Markov chain:

\begin{equation}
\begin{aligned}
P_t= \begin{cases}\exp \left(-\frac{\Delta E}{k_B T}\right), & \text { if } \Delta E>0 \\ 1, & \text { otherwise }\end{cases},
\end{aligned}
\end{equation}

\noindent where $\Delta E$ is the energy difference between the spin configurations, $k_B$ is the Boltzmann constant, and $T$ is the temperature of the system.
The hybrid mode has certain advantages, which we discuss in more detail in Section\,\ref{sec:practical-comparison}. 
In all MMCMC simulations in the present work, the temperature was set to 1\,$\mu$K.

We note that in \cite{Kwon2019} a neural-network-based method is described and used to identify the spin spiral state in a model system. However, that method contains neither any condition directing the global exploration of the PES, nor any perturbation scheme for $X$. In fact, the goal of the method in \cite{Kwon2019} is to identify the global minimum. That is orthogonal to our goal, which is to harvest a large number of metastable textures.

\subsection{Metastable spin texture isolation}
The local minima identified using the method described above usually correspond to several well separated spin textures within the simulation cell. Therefore, for practical reasons, each spin texture that we wish to analyze further needs to be isolated and embedded into a ferromagnetic background. Doing this "by hand" is cumbersome and time consuming. To solve the problem, we designed a pipeline that takes advantage of an AI-driven segmentation model. This approach greatly simplifies the process of harvesting the spin textures we want to analyze further. In the current pipeline, since we are working with a 2D system, we decided to use the state-of-the-art pre-trained Segment Anything Model (SAM), which is designed for image segmentation but can be easily transferred to work for our purposes, i.e., identifying complex spin textures in 2D spin systems, due to its zero-shot generalization property\,\cite{kirillov2023segment}. The workflow can be described as follows:
\begin{enumerate}    
  \item Translate the contents of the entire simulation cell (i.e, a normalized $3\times n$ matrix corresponding to a spin configuration) to an image by mapping the direction of each spin vector onto RGB color space.
  \item Generate spin texture masks using SAM.
  \item Isolate a selected spin texture with coordination information contained in the mask.
  \item Calculate the topological charge of the isolated spin texture and embed it into a ferromagnetic background.
  \item Store the embedded spin texture.
\end{enumerate}

\subsection{Computation of the spin Hamiltonian parameters and topological charges}
\label{sec:computation-parameters}

The electronic structure calculations of the fcc-Pd/Fe/Ir(111) multilayer system were performed using Density Functional Theory (DFT). We employed the self-consistent real-space linear muffin-tin orbital method in the atomic-sphere approximation (RS-LMTO-ASA) \cite{Peduto1991,Frota-Pessoa1992} with the local spin density approximation (LSDA) exchange-functional 
\cite{Barth1972}. The resulting ab-initio magnetic moments, as well as coupling parameters $J_{ij}$ and $\textbf{D}_{ij}$, used in Eq.\,\eqref{eq:param-hamiltonian}, are described in detail in \cite{miranda2022band}. To avoid effects coming from anticipated truncation, the coupling coefficients were considered up to 360 neighbors (within a distance of $\sim7a$ from the reference site, where $a=3.84\,$\AA$\,$is the experimental lattice parameter of the Ir host). Furthermore, the experimental value ($0.4$ meV/Fe \cite{Spethmann2022}) is here assumed for the $K^{\mathrm{U}}$ uniaxial anisotropy constant. 
Concerning the Zeeman term in Eq.\,\eqref{eq:param-hamiltonian}, we set the external field to be $B^{\mathrm{ext}}=3.5$\,T and applied in the out-of-plane direction ($[001]$). At this field, similarly to situation depicted in\,\cite{dupe2016}, the skyrmion lattice and ferromagnetic (single-domain) solutions are energetically almost degenerate \cite{miranda2022band}, being isolated skyrmions characterized as metastable structures. Finally, for the Gilbert damping parameter in SLLG, we used the value $\alpha\sim0.01$, which is in the order of magnitude of the intrinsic $\alpha$ in thin films (lower limit) \cite{Barati2014}. 

In the approximation where the Pd-induced spin moments are treated as independent degrees of freedom \cite{Polesya2010}, the Pd layer (when considered explicitly in the ASD simulations) reproduces the Fe layer spin texture \cite{miranda2022band}. Therefore, we restrict ourselves to the analysis of the Fe layer.

As it is widely known, skyrmionic structures are characterized by the topological number $Q$, which in the limit of a two-dimensional continuum reads:

\begin{equation}
Q=\frac{1}{4\pi}\int\mathbf{S}\cdot\left(\partial_{x}\mathbf{S}\times\partial_{y}\mathbf{S}\right)dxdy,
\end{equation}

\noindent that measures the winding number of $\mathbf{S}$. As we consider a discrete medium, the total topological charge, $Q_{\mathrm{tot}}$, of each spin configuration was calculated using the method proposed by Berg and Lüscher \cite{Berg1981}, with periodic boundary conditions. The quantity $Q_{\mathrm{tot}}$ 
accounts for all topological structures with individual topological charge (or winding number) $Q_i$ in the spin system ($Q_{\mathrm{tot}}=\sum_i Q_i$). Following the same reasoning, each of such structures can be isolated (e.g., in a ferromagnetic background) for the evaluation of the respective topological charges $Q_i$.

\subsection{Atomistic spin dynamics simulations}
To perform the stability analysis of the textures found by the algorithm presented in this work, we solve the stochastic Landau–Lifshitz–Gilbert (sLLG) equation \cite{UppASD_book}, which is implemented in the UppASD package, and is given as: 
\begin{equation}
\begin{aligned}\label{eq:sllg-eq}
\frac{d \mathbf{S}_i}{d t}=&-\gamma_{\mathrm{L}} \mathbf{S}_i \times\left(\mathbf{B}_i+\mathbf{B}_i^{\mathrm{f}}\right)\\
&-\gamma_{\mathrm{L}}\alpha\mathbf{S}_i \times\left[\mathbf{S}_i \times\left(\mathbf{B}_i+\mathbf{B}_i^{\mathrm{f}}\right)\right].
\end{aligned}
\end{equation}
%
Here, 
$\mathbf{B}_i = -\frac{1}{\mu_i}\frac{\partial\mathscr{H}}{\partial\mathbf{S}_i}$ is the effective field on site $i$, related to the Hamiltonian defined in 
Eq.\,\eqref{eq:param-hamiltonian} and parametrized by the ab-initio quantities described in Section \ref{sec:computation-parameters}. The dimensionless (and isotropic) Gilbert damping parameter is here denoted by $\alpha$, while $\gamma_{\mathrm{L}}=
\gamma / \left(1+\alpha^2\right)$ 
is the renormalized gyromagnetic ratio (as a function of the bare one, $\gamma$). In turn, $\mathbf{B}_i^{\mathrm{f}}$ is the stochastic field arising from the thermal fluctuations. This term describes how temperature effects are considered in the spin dynamics, in a Langevin approach \cite{skubic2008method}. More specifically, $\mathbf{B}_i^{\mathrm{f}}$ is modelled by an uncorrelated Gaussian white noise, so that the time-averages $\left<\mathbf{B}_i^{\mathrm{f}}(t)\right>=0$ and $\left<B_{i\mu}^{\mathrm{f}}(t)B_{j\nu}^{\mathrm{f}}(t^{\prime})\right>=2D\delta_{ij}\delta_{\mu\nu}\delta(t-t^{\prime})$ for the site indices $i$ and $j$ and the Cartesian coordinates $\mu,\nu=\{x,y,z\}$. In the second relation, the parameter $D$ can be obtained via the fluctuation-dissipation theorem \cite{Mentink2010} as $D=\alpha{k_{B}T}/{(\gamma \mu_i)}$, for a given temperature $T$, where $k_B$ is the Boltzmann constant.

In Eq.\,\eqref{eq:sllg-eq}, the first term expresses the precessional motion of atomistic magnetic moments in a given system, and the second term describes the damped motion (both in the adiabatic approximation, i.e., ${d\left|\mathbf{S}_{i}\right|}/{dt}=0$). Besides the dynamics, in the absence of thermal effects, the solution of Eq.\,\eqref{eq:sllg-eq} with $\alpha>0$ corresponds to the relaxation process of finding the nearest energy minimum in the PES.

Throughout the text, we define the typical simulation times of 200 ps (for the stability of the topological charge) and 2 ns (for the calculation of the average energy/spin with $T>10^{-4}$ K). The numerical solution of Eq.\,\eqref{eq:sllg-eq} is obtained with a time step of $\Delta t=10^{-16}$ s.

\begin{figure*}
\centering
\includegraphics[width=16cm]{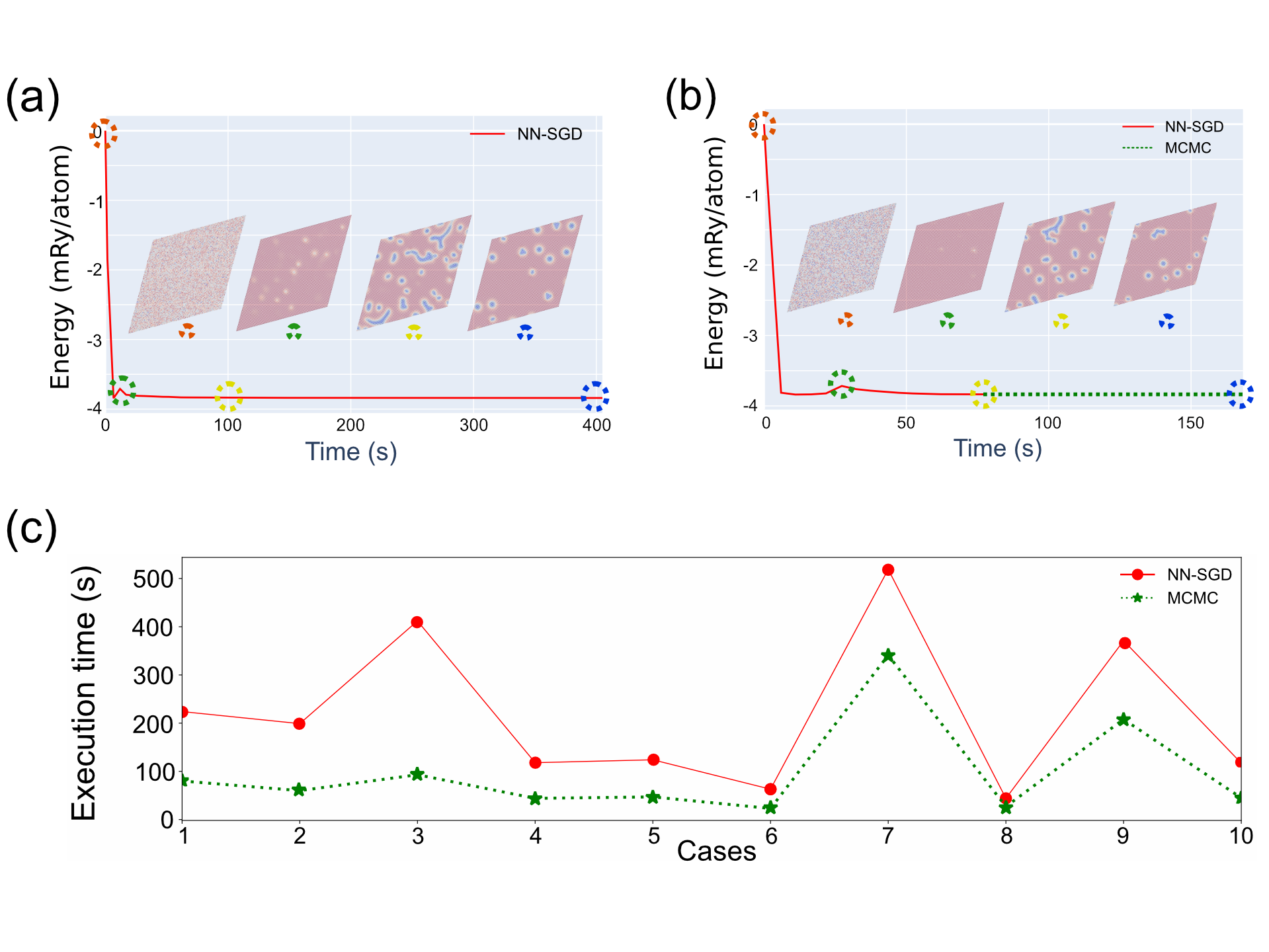}
\caption{Illustration of (a)
SGD-only mode (NN-SGD), (b) hybrid mode (NN-SGD followed by MMCMC), and (c) comparison of the execution times of the two modes from the user perspective. Panels (a) and (b) show schematically the optimization procedures for the two modes. Note that the two optimizations in (a) and (b) start from different (random) spin configurations, and cannot be compared directly. The red dashed circle is the starting point (a random spin texture). At the green dashed circle, a spin configuration that does not fulfill Eq.\,\eqref{HPTP}) is encountered. Therefore, the algorithm creates a new starting point by adding Gaussian noise to $X$, and restarts the optimization toward another local minimum.
The yellow dashed circle indicates the point at which the intermediate convergence criterion is fulfilled. At this point, the hybrid mode switches from using NN-SGD to MMCMC.
Finally, the blue dashed circle points out the final converged solution. The parallelogram insets in both plots (a/b) show the magnetic textures of the system at different times during the minimization procedure. Illustrative movies of the optimization processes can be found in the Supplementary Material.
 Panel (c) shows the execution times of both algorithms, for ten randomly selected cases. The NN-SGD optimization was performed on a single Nvidia A100 GPU, whereas the MMCMC optimization used an Intel Xeon Gold 6130 CPU, parallelized with OpenMP. 
   }
    \label{fig:texturemcml}
\end{figure*}
\subsection{Comparison of the SGD-only and hybrid modes}
\label{sec:practical-comparison}

From the user perspective, it is important that the computations converge quickly.
Although we cannot directly compare the efficiency of the two optimization algorithms (NN-SGD and MMCMC) since, among other reasons, practically, in our implementation they use different architectures (GPU and CPU parallelized by OpenMP, respectively), and are written in different programming languages, we can compare the execution times of the SGD-only and hybrid modes from the user perspective. 

To achieve this practical comparison, we introduce an intermediate convergence hyperparameter $T_{s}^{\textnormal{int}}$ in the context of 
Eq.\,\eqref{eq:Ts}. When the optimization has reached the intermediate level of precision defined by
$T_{s}^{\textnormal{int}}$,
we either continue with the NN-SGD optimizer (the SGD-only mode), or switch to the MMCMC optimizer (the hybrid mode) until the stopping condition Eq.\,\eqref{eq:Ts} is fulfilled for our final convergence hyperparameter $T_s^{\textnormal{final}}$.
In the present benchmark, we have used 
$T_{s}^{\textnormal{int}}=10^{-3}$\,mRy/atom and 
$T_s^{\textnormal{final}}=10^{-6}$\,mRy/atom.
For both modes, the magnetic texture eventually converges to the same configuration on the PES, but with distinct execution times.  Typical optimization procedures for the SGD-only and hybrid modes, for different initial spin configurations, are illustrated in Fig.~\ref{fig:texturemcml}(a) and (b), respectively. 

In Fig.~\ref{fig:texturemcml} (c), we compare the execution times of the two modes for ten randomly selected initial spin configurations.
In order to produce a fair comparison, both algorithms (NN-SGD and MMCMC) start at $T_s^{\textnormal{int}}$ from the same point within each case, and the minimization procedure is stopped at $T_s^{\textnormal{final}}$.
The red (green) dots show the execution times, counted from the intermediate threshold, for the SGD-only (hybrid) mode. Clearly, the hybrid scheme shows consistently shorter execution times from the user perspective.

\section{Results}
\label{results}
%
\begin{figure*}
    \centering
    \includegraphics[width=16cm]{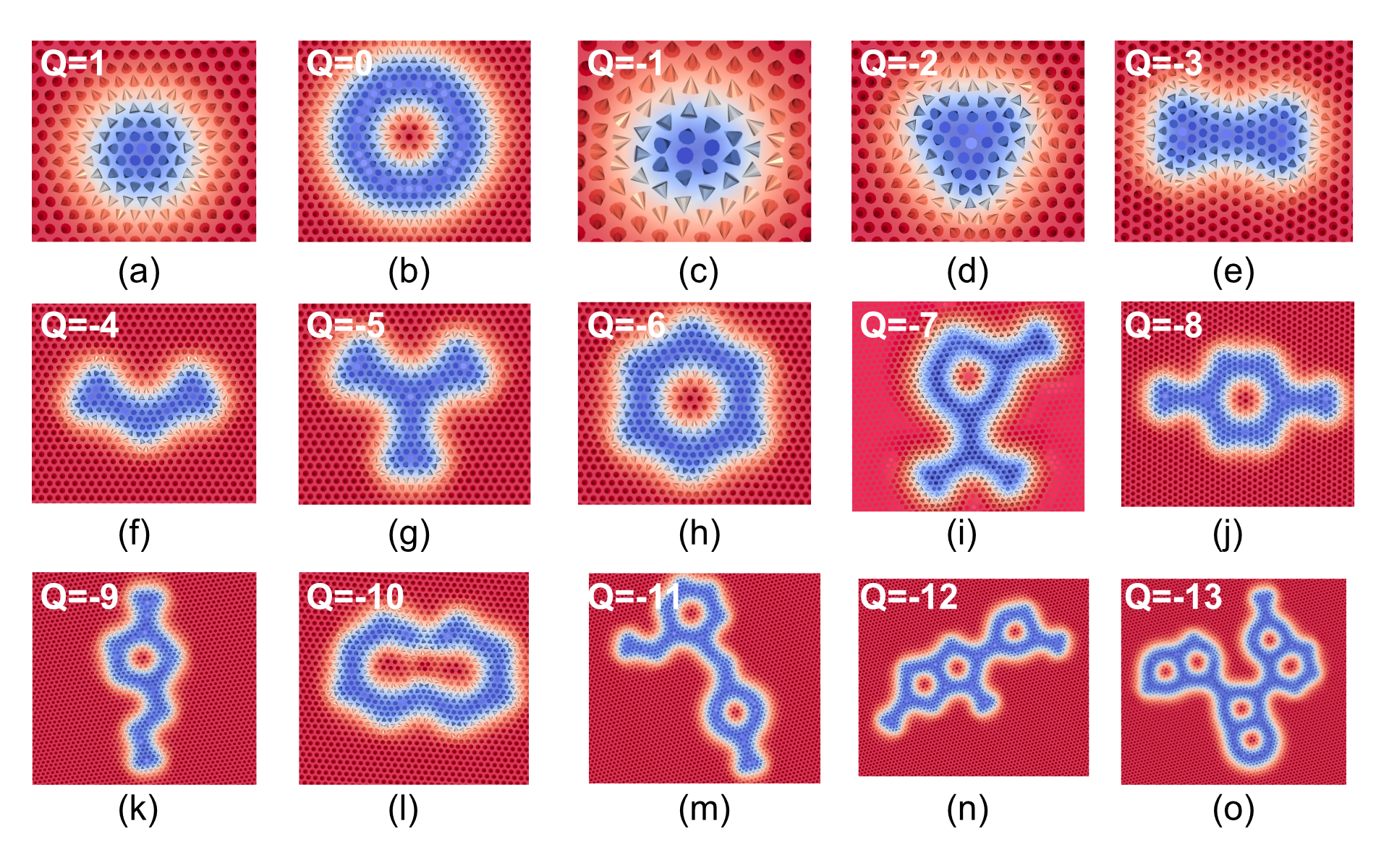}
    \caption{A subset of identified topological metastable spin textures in the Pd/Fe/Ir(111) system with $B^{\mathrm{ext}} = 3.5$\,T. (a) skyrmion, (b) skyrmionium, (c) antiskyrmion, (d) to (o): isolated magnetic textures, with higher-order topological charge varying between  $Q=-2$ and $Q=-13$. The distance between neighboring spins (cones) is $a\frac{\sqrt{2}}{2}\sim2.72\,$\AA. Blue and red colors indicate opposite spin directions.} 
    \label{fig:texture3}
\end{figure*}
\subsection{Textures with higher-order $Q$}
We applied our algorithm in the hybrid mode to systematically harvest topological spin textures in the fcc-Pd/Fe/Ir(111) system. 
Overall, given the right conditions such as for instance an appropriate magnitude of the applied magnetic field, we find that this system can hold many types of spin textures including skyrmions, skyrmionium, antiskyrmions, rings and bags comparable to the ones obtained in other systems~\cite{PhysRevLett.117.157205, PhysRevB.95.094423,rybakov2019chiral, Foster2019, PhysRevB.102.144422,dupe2016}.
For the remainder of this work, we have selected to focus our analysis on  antiskyrmion textures. We found an abundance of such textures, and, curiously, they have been less discussed in the literature compared to, e.g., skyrmion bags.
In Fig.~\ref{fig:texture3}, we show 15 selected spin textures that have been identified to be metastable by our algorithm, with the topological charge $Q$ varying from $1$ to $-13$. Note that throughout this work we use the convention that the skyrmion has topological charge $+1$, and antiskyrmions have negative topological charge. 
The first row in Fig.~\ref{fig:texture3} shows a $Q=1$ skyrmion, skyrmionium ($Q=0$), an three simpler antiskyrmions. As $|Q|$ increases, the number of limbs and/or rings in the textures increase, see rows 2 and 3 in Fig.~\ref{fig:texture3}. For large $|Q|$, we find very convoluted textures with several rings and limbs combined.

\subsection{Stability analysis and binding energies} 

\begin{figure}[!h]
    \centering
    \includegraphics[width=8cm]{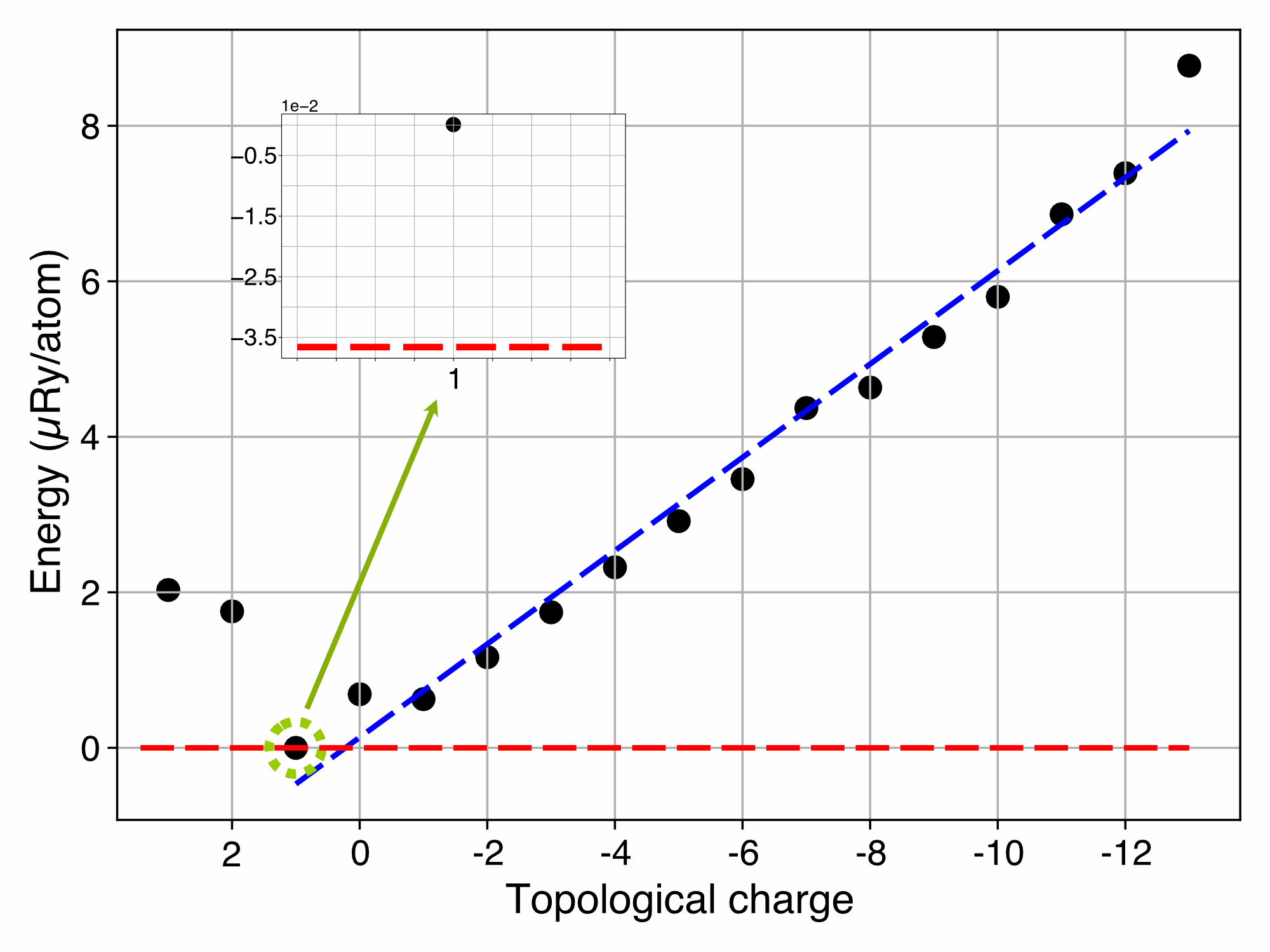}
    \caption{Relative energy per atom of isolated spin textures with topological charge $Q$ ranging from $-13$ to 1. (For completeness we have also included the $Q=+2$ and $Q=+3$ cases in the plot).  The total energy is calculated by embedding each topological excited state in an fcc-Pd/Fe/Ir(111) supercell i.e., a single-domain ferromagnetic background with size $23.6\times27.2$ nm$^2$. The energy scale in the figure is set so that the lowest energy (i.e., the $Q=1$ skyrmion state) is zero, for visibility. The black dots are computed energies and the blue dashed line is a linear fit to the data. The horizontal 
   red dashed line shows the energy of the ferromagnetic state. The inset shows the energy gap between the $Q = -1$ spin texture and the ferromagnetic state.
    }
    \label{fig:energy}
\end{figure}

\begin{figure*}
    \centering
    \includegraphics[width=16cm]{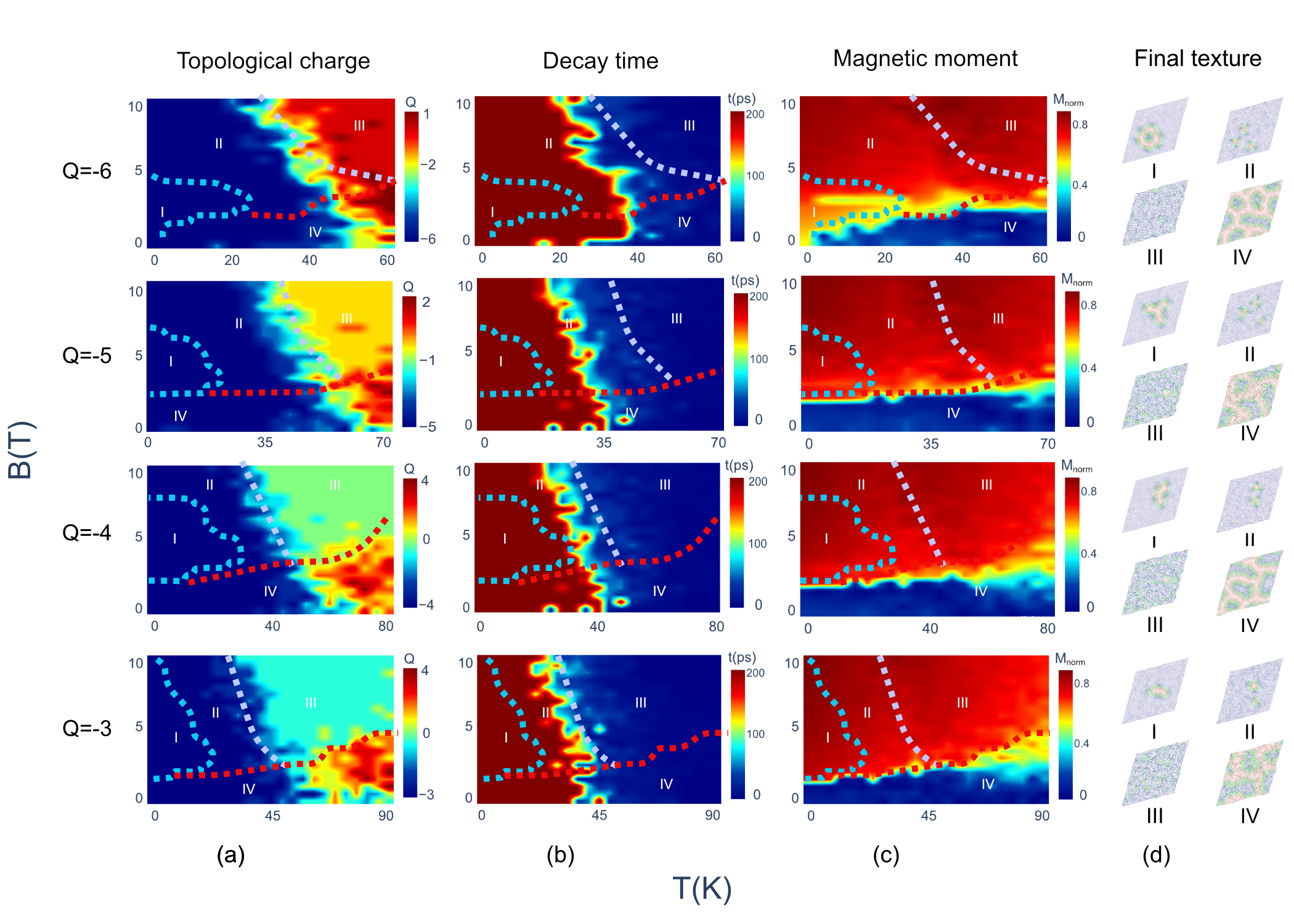}
    \caption {Magnetic field versus temperature stability maps of antiskyrmions with topological charge varying from $Q=-6$ up to $Q=-3$. Columns 1 to 3 show the final topological charge, topological charge decay time and the normalized magnitude of the final magnetic moment, respectively. The top row is for $Q = -6$, the second row is for $Q = -5$, the third row is for $Q = -4$ and the bottom row is for  $Q = -3$. In the right-most column we show typical final magnetic textures in each 
   zone (I, II, III and IV). 
   }
    \label{fig:phase_diagram}
\end{figure*}

Generally speaking, for a given topological charge $Q$, several spin textures are typically possible, with varying number of holes and chiral kinks. Here, holes are defined as non-self-intersecting curves that define closed domain walls in the structure. Chiral kinks are described in more detail in  \cite{PhysRevB.102.144422}. 
To elucidate how the stability of the spin textures vary with $Q$, we have for each $Q$ selected the lowest-energy spin texture found in our simulations. 
In Fig.~\ref{fig:energy}, we show the computed energy difference with respect to the  
ferromagnetic single-domain state ($Q=0$), $\Delta E=E-E_{\textnormal{FM}}$ for these selected spin textures. We see that the energy difference, $\Delta E$, is almost linear with respect to $Q$.
As $Q$ decreases one unit, the energy per spin increases on average $\sim0.6$ $\mu$Ry. 
Consequently, the antiskyrmion states with $\left|Q\right|\geq1$ are relatively close in energy.

\begin{figure}[!h]
    \centering
    \includegraphics[width=8cm]{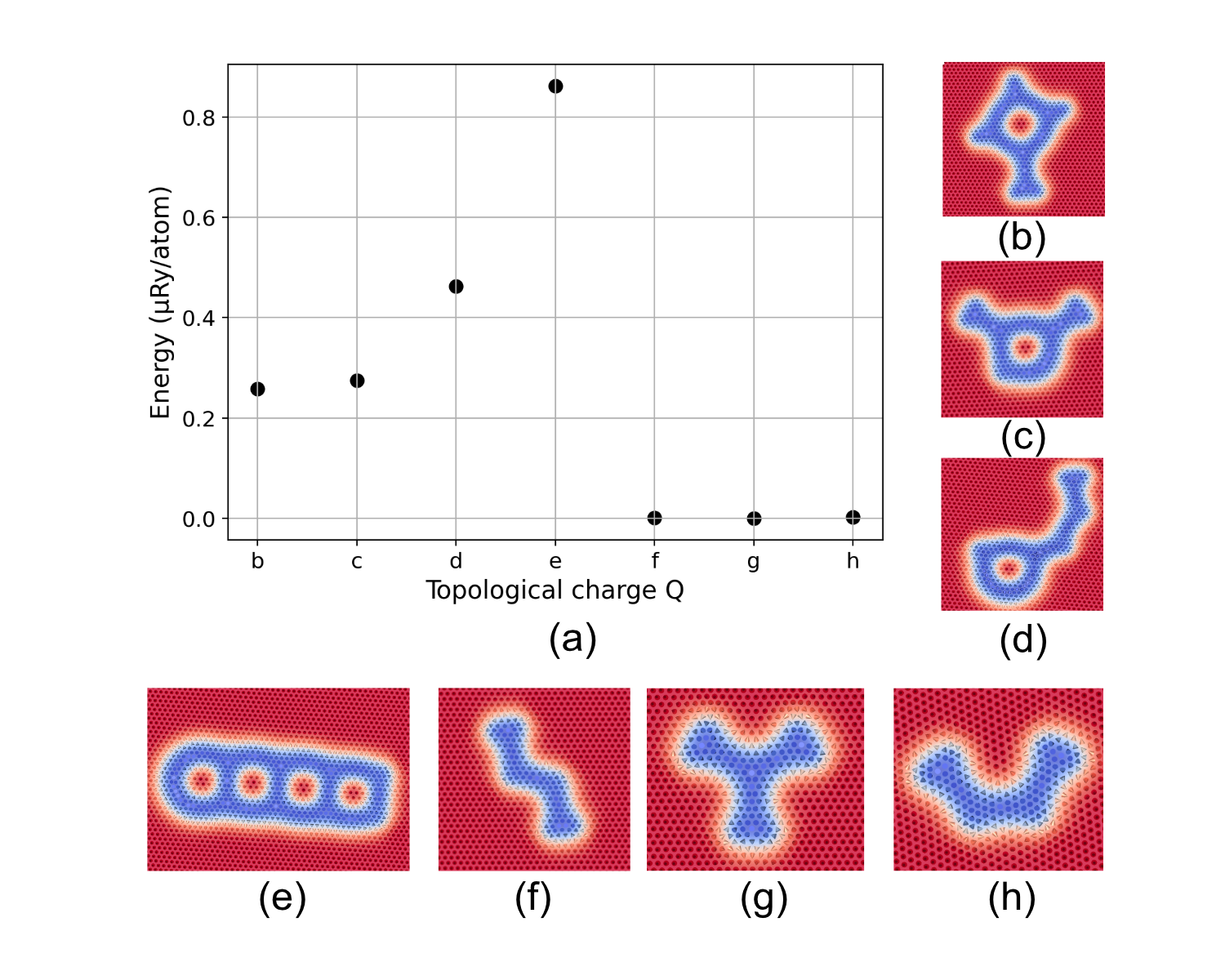}
    \caption{(a) Relative energy, with respect to the ferromagnetic single-domain state, per atom of the isolated topological textures with topological charge $Q=-5$ in a magnetic field of 3.5\,T, shown in panels (b) to (h). The energy scale here is set so that the energy is zero for the texture with the lowest energy per atom. }
    \label{fig:texture_time}
\end{figure}

To gain a more detailed understanding of the stability and lifetimes of these topological spin textures, we have analyzed them further using the concept of skyrmionic binding energy $E_b(Q)$ developed in \cite{PhysRevB.95.094423}, in combination with results from atomistic spin dynamics simulations using the UppASD software package \cite{UppASD_book}.
In\cite{PhysRevB.95.094423}, the skyrmion binding energy $E_b(Q)$  is defined as (respecting our $Q$ sign convention)
\begin{equation}\label{eq:binding-energy}
E_b(Q) = E(Q) - \left|Q\right|E(\textnormal{sign}\, Q) - mE(Q=0),
\end{equation}
\noindent where $E(\textnormal{sign}\, Q)= \begin{cases}E(Q=-1), & \text { if } Q < 0 \\ E(Q=1), & \text{ if } Q > 0\end{cases}$, and $m$ is the number of $Q=0$ structures that compose the complex skyrmionic configurations. \\ \\

Here, an underlying assumption is that the higher-order spin textures  can be described as an excited state consisting of multiple bounded $Q=\pm1$ (and eventually $Q=0$) structures.
Using this definition, in Table \ref{tab:binding-energies} we show the computed binding energies of higher-order antiskyrmions 
with $Q\in[-5,-3]$, where $m=0$. The negative $E_b$'s obtained for these antiskyrmions  mean that it is energetically favorable for lower-$|Q|$ structures to combine together and form these more complex structures with higher total $|Q|$. 

For comparison, we also computed the binding energy of a $Q = 2$ texture, in which two skyrmions are trapped inside skyrmionium, i.e., a skyrmion bag, see the last row in Table \ref{tab:binding-energies}.
Note that for this case, Eq.\,\eqref{eq:binding-energy} must account also for the skyrmionium energy, so that $m=1$.
We find that the binding energy of this skyrmion bag is positive. Similarly, skyrmions bags with $Q> 2$ were also found to have positive binding energies. Thus, the given $(B,T)$ conditions disfavor lower-$Q$ structures to generate more complex structures with a higher positive-$Q$.

\begin{table}[t]
\caption{\label{tab:binding-energies}%
Energies $E(Q)$ of the topological structures with respect to the ferromagnetic (single-domain) background, and binding energies $E_b(Q)$ of higher-order antiskyrmions (see Eq.\,\eqref{eq:binding-energy}), both given in $\mu$Ry/atom. The out-of-plane applied field is $B=3.5$ T, at the temperatures indicated in parenthesis.}
\begin{ruledtabular}
\begin{tabular}{ccc}
$Q$ & $E(Q)$ & $E_b(Q)$ \\ \\
\multirow{3}{*}{$+1$} & $0.04$ ($<10^{-4}$ K) & \multirow{3}{*}{n.d.\footnote{n.d. = not defined}} \\
 & $0.04$ ($3$ K) & \\
 & $0.04$ ($20$ K) & \\ \\
\multirow{3}{*}{$-1$} & 0.67 ($<10^{-4}$ K) & \multirow{3}{*}{n.d.} \\
 & $0.67$ ($3$ K) & \\
 & $0.67$ ($20$ K) & \\ \\
\multirow{3}{*}{$-3$} & 1.78 ($<10^{-4}$ K) & $-0.22$ \\
 & $1.79$ ($3$ K) & $-0.21$ \\
 & $1.80$ ($20$ K) & $-0.20$ \\ \\

\multirow{3}{*}{$-4$} & 2.36 ($<10^{-4}$ K) & $-0.31$ \\
 & $2.36$ ($3$ K) & $-0.30$ \\
 & $2.42$ ($20$ K) & $-0.25$ \\ \\
\multirow{3}{*}{$-5$} & 2.95 ($<10^{-4}$ K) & $-0.39$ \\
 & $2.95$ ($3$ K) & $-0.37$ \\
 & $3.02$ ($20$ K) & $-0.31$ \\ \\
 \multirow{2}{*}{$+2$} & 1.76 ($<10^{-4}$ K) & $0.94$ \\
 & $1.76$ ($3$ K) & $0.95$ 
\end{tabular}
\end{ruledtabular}
\end{table}

Figure \ref{fig:phase_diagram} depicts the results from our spin-dynamics simulations for selected topological spin textures with a $Q$ ranging from $-3$ to $-6$. Specifically, we have mapped out how the topological charge, decay time $\tau$, and total magnetic moment changes with temperature and applied magnetic field.
Here, the decay time $\tau$ is the time it takes for the spin texture to unpack, i.e., change its topological charge.
Based on the decay time, final topological charges, and final magnetic moment (per site), and also visual inspection, we divided the charts into four basic zones. 
Thus, zone I indicates the region in field and temperature in which the topological particle-like structures are long-lived (here taken to mean a life time exceeding $\tau>200$ ps), keeping their initial real-space structure intact. 
When $B$ and/or $T$ become sufficiently high to reduce the binding energy (see Eq.\,\eqref{eq:binding-energy}) by favoring the FM state or due to thermal fluctuations, these higher-order antiskyrmions unpack into lower-$|Q|$ textures. 
These lower-$|Q|$ structures can either be relatively stable or survive for just a few picoseconds, depending on the energy barrier height. 
This characterizes the transition region (zone II), where the initial higher-order antiskyrmions are unstable but topological particle-like structures are still present. With further increased temperature, even the lower-$|Q|$ textures become unstable, gradually shifting to a mostly FM spin configuration (note that here $T \lesssim T_{C}$ \cite{miranda2022band}), which defines zone III. Finally, in zone IV we find metastable configurations that resemble domain structures. In that zone, the sufficiently low $B$ either drives the spin-spiral to be the dominant state, 
or there is a mix of a spin-spiral state and a skyrmion lattice. As a general trend, we notice that the region where the higher-order antiskyrmions are long-lived (zone I) diminishes with increasing initial $\left|Q\right|$.
These observed patterns agree well with the information in Table \ref{tab:binding-energies}. $E_b$ increases with temperature, enhancing also the probability that high-$|Q|$ structures separate into lower-$|Q|$ (or 
$Q=-1$) constituents due to thermal effects.

The above analysis, together with the facts that (\textit{i}) single antiskyrmions are long-lived up to sizable temperatures ($T\sim 32$ K at $B^{\mathrm{ext}}=3.5$ T), which makes the population of these metastable states more likely (with a finite probability given by the Boltzmann factor); and (\textit{ii}) higher-order antiskyrmions are relatively close in energy (Fig. \ref{fig:energy}), suggests that some high-$Q$ states in Pd/Fe/Ir(111) can be formed experimentally, at their critical temperatures, by merging $Q=-1$ textures, as similarly proposed in \cite{Ritzman2020}. The process of bound states generation, however, must overcome an energy barrier.

\subsection{Geometry-topology analysis}
Our method is able to harvest numerous metastable topological spin textures. This opens the door for analyzing whole families of spin textures with the same $Q$. It is for instance interesting to gain insight into how the stability is affected by the internal structure of spin textures with a given $Q$.  
In Fig.~\ref{fig:texture_time}, the relative energies of seven spin textures, all with $Q=-5$, are plotted.
We see that the spin textures have different energies even though all of them have the same topological charge. 
This also implies that Eq.\eqref{eq:binding-energy} is indeed  a simplification.
The structures can be classified into three different groups. The textures shown in Figs.~\ref{fig:texture_time} (b)-(d) can all be described as a skyrmion ring (or hole) with limbs of varying number and lengths attached, whereas the texture in Fig.~\ref{fig:texture_time} (e) consists of several rings directly attached to each other. 
The remaining textures (Figs.~\ref{fig:texture_time}(f)-(h)) represent hole-free antiskyrmions. Overall, the textures with holes have higher energy and are, consequently, less stable than the ones without holes inside. This result agrees well with what can be observed from Figs.~\ref{fig:texture3} and~\ref{fig:energy}, i.e., spin textures with holes 
tend to be less stable as compared to the textures without holes.
We stress that the textures in Fig.~\ref{fig:texture_time} do not cover the whole diversity of solutions for $Q=-5$ (some other possible metastable states with $Q=-5$ are given in Ref. \cite{PhysRevB.102.144422}).

\section{Discussion and conclusions}
\label{conclusion}
Several factors may influence the results obtained with the computational approach we have presented and used in this work.
Which metastable spin textures (i.e., local minima) our method finds of course depends on which starting points we allow for the optimization. Thus, tuning the condition in Eq.\,\eqref{HPTP} will strongly affect the results. In this work, we have selected to impose a simple condition on the DM interaction strength, i.e., $T_{\mathrm C}^{\textnormal{low}}$  > 0.01. Many other options are possible, including imposing several different conditions simultaneously. A condition that would select on exchange frustration may tip the balance in favor of other types of local minima\,\cite{PhysRevLett.117.157205,Rozsa2017}.

An important issue is the shape of the energy barriers surrounding each local minimum. The probability for a starting point $X$ to end up on a wider energy barrier is higher than for it to end up on a more narrow energy barrier. Thus, our method will naturally select for local minima with wide (but perhaps low) energy barriers, unless this is remedied by imposing a condition to specifically select for narrow energy barriers.

In addition, the stability of the high-order topological spin textures may depend quite sensitively on the parameters used in the spin Hamiltonian. As an example, we mention that our $\left|Q\right|>6$ textures have a sensitive dependence on the exchange frustration. In our case, the major source of frustration is the next-nearest neighbor interaction shell, so that the relation $J_3/J_1$ (see Appendix \ref{sec:j-fe-fe-appendix}) -- where $J_n$ denotes the Fe-Fe interaction from a reference site to its $n$-th neighboring shell -- plays a significant role. Indeed, a change of $\sim10\%$ in $J_3/J_1$ can lift the theoretical metastability of those configurations. In the context of stability analysis, it is worth  noting that there exist additional higher-order interactions that have not been taken into account in this study. These higher-order interactions can influence the energy barrier height as has been demonstrated in recent investigations on skyrmions \cite{Paul2020,Gutzeit2021}.
Optimizing the condition\,\eqref{HPTP} for various scenarios and performing additional sensitivity analysis will be the subject of future work. 

To conclude, we have presented a new computational approach based on a conditional neural network, aimed at identifying large numbers of higher-order topological spin textures. 
Using the Pd/Fe/Ir(111) model system as an example, we have harvested numerous hitherto unidentified antiskyrmion spin textures. 
The identified spin textures have in turn enabled us to perform a systematic analysis of how the relative stability of spin textures depends on temperature, external magnetic field, the topological charge as well as the number of holes in the texture.

\section{Data availability}
All data is available from the authors upon request.

\section{Code availability}
All the relevant code is available from the authors upon request.

\begin{acknowledgments}
This work was financially supported by the Knut and Alice Wallenberg Foundation (grant numbers 2018.0060, 2021.0246, and 2022.0108), 
Vetenskapsrådet (grant numbers 2017-03832, 2019-03666, 2016-05980, and 2019-05304), 

the European Research Council (grant number 854843-FASTCORR), the foundation for Strategic Research SSF, and China Science Council (CSC) grant number 201906920083.
Support from STandUP and eSSENCE is also acknowledged.

Computations/data handling were enabled by
resources provided by KAW (Berzelius-2022-259) and the Swedish National Infrastructure for Computing (SNIC), partially funded
by the Swedish Research Council through grant agreement No. 2018-05973.

The authors also acknowledge discussions with Zhuanglin Shen.
\end{acknowledgments}

\bibliography{Qichen_Antiskyrmion}

\clearpage

\section{Supplementary}

\subsection{Minimum energy paths}
\label{sec:mep-appendix}
\begin{figure*}
    \centering
    \includegraphics[width=16cm]{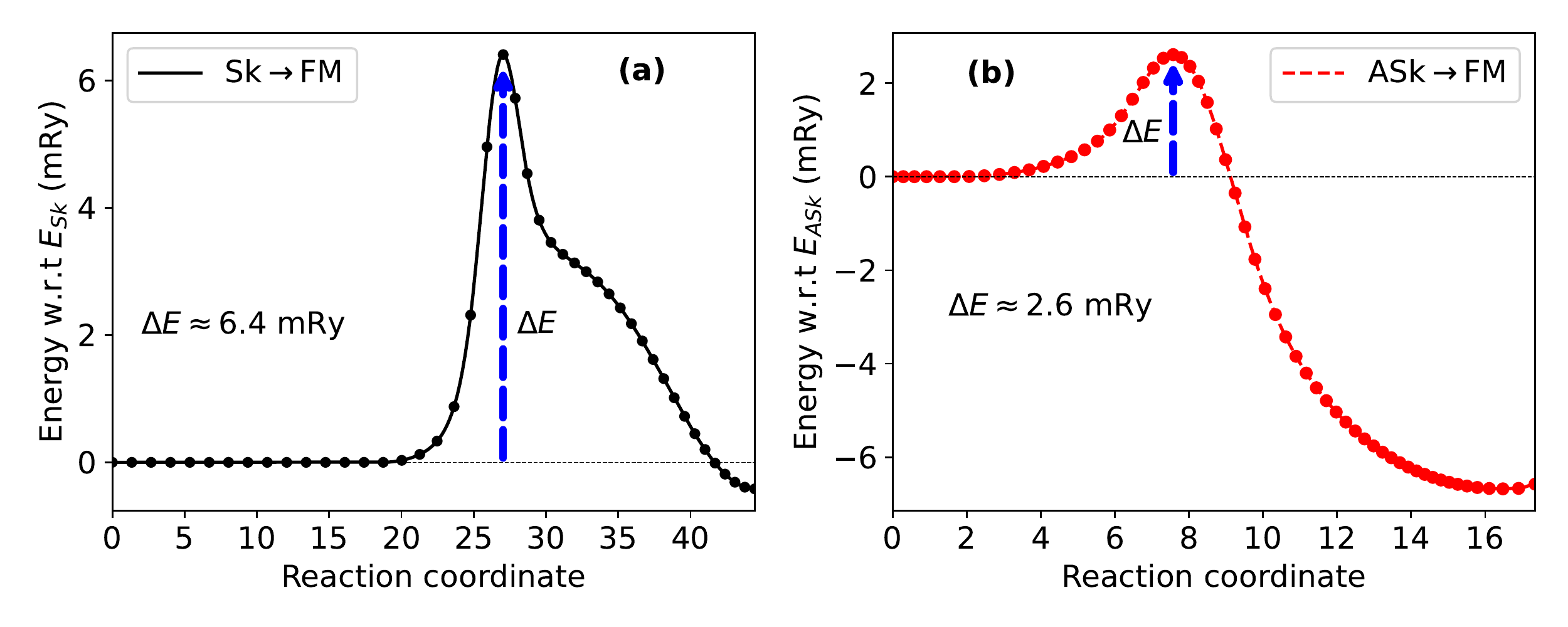}
    \caption {Minimum energy paths to collapse, at $T=10^{-4}$ K and $B^{\mathrm{ext}}=3.5$ T, for: (a) single skyrmion; (b) single antiskyrmion on Pd/Fe/Ir(111). The energies are given with respect to the initial state (i.e., isolated $Q=\pm 1$ structure) and mapped as a function of the reaction coordinate corresponding to the collapse path to the ferromagnetic (FM) single-domain state. The maximum energy represents a first order saddle point in the potential energy surface.}
    \label{fig:gneb-curves}
\end{figure*}

The calculation of the minimum energy paths (MEPs) between two local minima in the energy surface results in the corresponding activation barrier. Considering one of the states as the ferromagnetic (single-domain) configuration, then we can define the so-called path to collapse, and the resulting curve can be directly related to the stability of the topological texture. 
Here, the geodesic nudged elastic band (GNEB) method \cite{Bessarab2015}, as implemented in UppASD, is used in order to determine the MEP to collapse of a single skyrmion (or antiskyrmion) in Pd/Fe/Ir(111), considering the configuration space determined by the parameters described in Ref. \cite{miranda2022band}. The resulting MEPs are shown in Fig.\,\ref{fig:gneb-curves}, where the highest activation barrier is obtained for the isolated skyrmion. This is consistent with previous calculations (e.g., Ref. \cite{Malottki2017}), and demonstrates the enhanced stability of skyrmions in the context of annihilation to the ferromagnetic state. 

\subsection{Intra-layer interactions in Pd/Fe/Ir(111)}
\label{sec:j-fe-fe-appendix}
\begin{figure*}
    \centering
    \includegraphics[width=12cm]{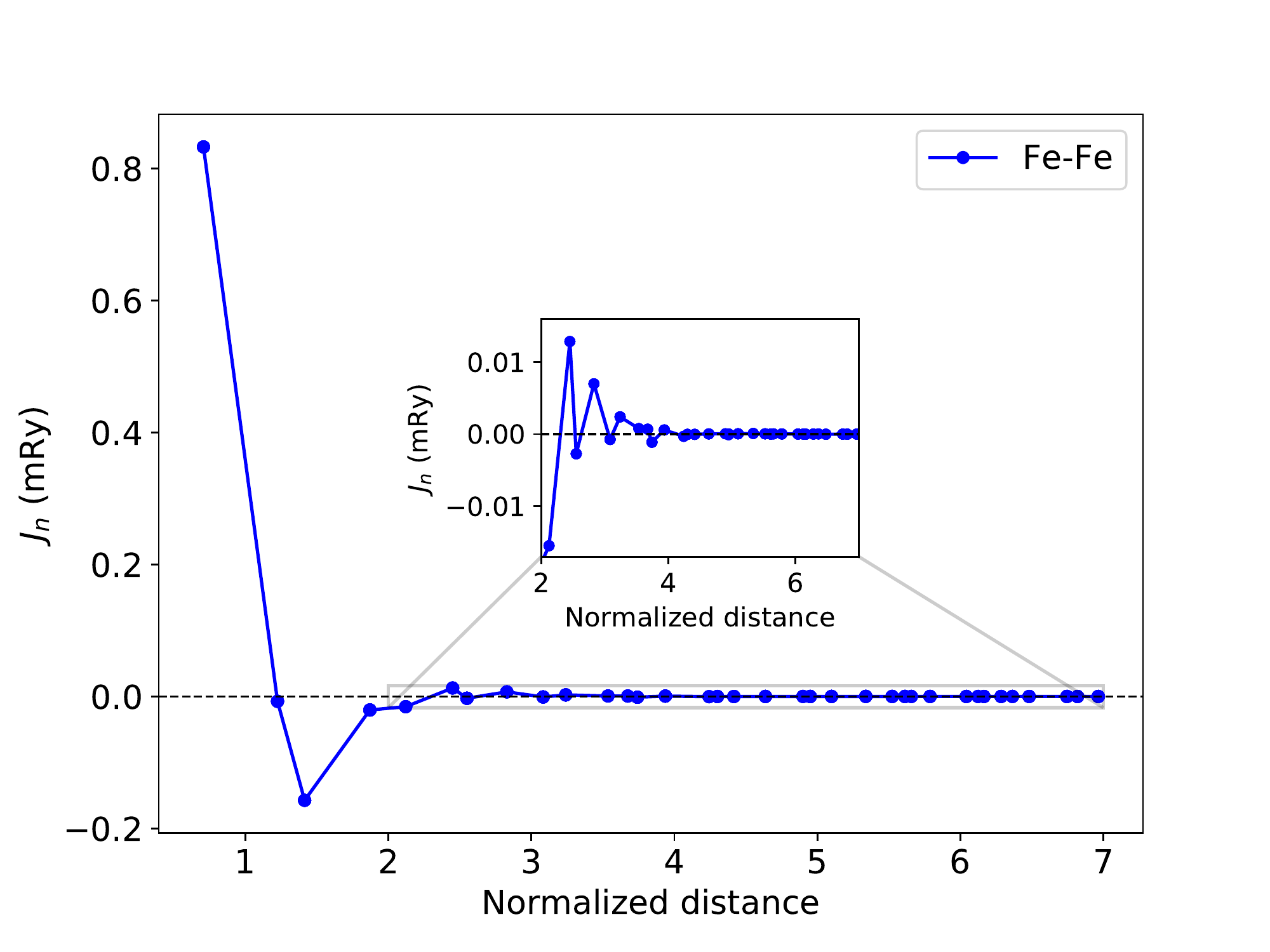}
    \caption {The Heisenberg Fe-Fe exchange interactions as a function of the (normalized) distance in Pd/Fe/Ir(111).}
    \label{fig:j-fe-fe}
\end{figure*}

Fig.\,\ref{fig:j-fe-fe} shows the isotropic Heisenberg exchange interactions of a reference Fe site with its surrounding Fe neighbors, calculated using the magnetic force theorem \cite{Liechtenstein1987} and analyzed in detail in\,\cite{miranda2022band}. Here, $J_n$ denotes the interaction between the reference site and the $n$-th neighboring shell.

\subsection{Supporting movies}
Examples of the energy minimization process and the resulting metastable state of Pd/Fe/Ir(111) containing high-order antiskyrmions, obtained with the use of the SGD-only (hybrid) mode, can be found in the movies S1 (S2). In the movies, the frames in red and blue colors refer to optimization with NN-SGD, and the gray and yellow frames refer to optimization using MMCMC, respectively. The red/gray colors represent the out-of-plane component of the magnetic moments, while the blue/yellow colors represent the in-plane components.

\end{document}